\documentclass{llncs}
\usepackage{wrapfig}
\usepackage{llncsdoc}
\usepackage{epsfig,psfrag}
\usepackage{latexsym,amsfonts,amssymb,amsmath}
\usepackage{longtable}
\usepackage{tikz,pgf}
\usepackage{subcaption}
\usepackage[export]{adjustbox}
\usepackage{rotating}
\usepackage{bm}
\usepackage[normalem]{ulem}
\usepackage{comment} 
\usepackage{tikz}
\usetikzlibrary{patterns,positioning,arrows,decorations.markings,shadows}
\usepackage{pgfplots}
\usepackage{listings}
\usepackage{graphicx}
\usepackage{epstopdf}
\usepackage{color}
\usepackage{float}
\usepackage{lscape}
\usepackage{array}
\usepackage{amsmath,amsfonts}
\usepackage{booktabs}
\usepackage{blkarray}
\graphicspath{ {./graphs/} }
\usepackage{multirow}
\usepackage{color, colortbl}
\definecolor{Gray}{gray}{0.9}

\makeatletter
\renewcommand*\env@matrix[1][c]{\hskip -\arraycolsep
	\let\@ifnextchar\new@ifnextchar
	\array{*\c@MaxMatrixCols #1}}
\makeatother

\newcommand{\bqn}{\begin{eqnarray}}
\newcommand{\eqn}{\end{eqnarray}}
\newcommand{\bq}{\begin{eqnarray*}}
\newcommand{\eq}{\end{eqnarray*}}

\usepackage{url}
\usepackage{hyperref}
\hypersetup{backref,colorlinks,linktocpage=true}

\DeclareMathOperator*{\argmin}{arg\,min}
\DeclareMathSymbol{\sminus}{\mathbin}{AMSa}{"39}

\begin{document}

\title{Spanning Tree Basis for Unbiased Averaging of Network Topologies}

\author{
Sixtus Dakurah\inst{1}
}

\institute{Department of Statistics, University of Wisconsin-Madison\\
\textit{sdakurah@wisc.edu}
}
\maketitle

\begin{abstract}
In recent years there has been a paradigm shift from the study of local task-related activation to the organization and functioning of large-scale functional and structural brain networks. However, a long-standing challenge in this large-scale brain network analysis is how to compare network organizations irrespective of their complexity. The maximum spanning tree (MST) has served as a simple, unbiased, standardized representation of complex brain networks and effectively addressed this long-standing challenge. This tree representation, however, has been limited to individual networks. Group-level trees are always constructed from the average network or through a bootstrap procedure. Constructing the group-level tree from the average network introduces bias from individual subjects with outlying connectivities. The bootstrap method can be computationally prohibitive if a good approximation is desired. To address these issues, we propose a novel spectral representation of trees using the \textit{spanning tree basis}. This spectral representation enables us to compute the average MST and demonstrate that this average tree captures the global properties of all the MSTs in the group and also overlaps with the union of the shortest paths in the functional brain networks.
\end{abstract}

\section{Introduction}

This work addresses a central problem in network neuroscience, developing unbiased and computationally efficient methods for representing group level brain network topologies, with a particular focus on functional brain networks.
The human brain is a complex network of functional and structurally connected brain regions. Understanding the integration of these regions, therefore, requires a good knowledge of the underlying topology of the brain \cite{torres2021and,dakurah2022modelling,dakurah2025discrete,dakurahregistration,dakurah2025brain}. This often involves studying the network connectivity pattern, which reveals important information about the human brain's structural, functional, and causal organization. Among the many techniques for studying this connectivity pattern, functional connectivity has been the major focus of computational studies in recent years \cite{bullmore2009complex,goldenberg2015use,songdechakraiwut2020dynamic,huang2019dynamic}. Functional connectivity revolves around the temporal correlations between neurophysiological events at different spatial resolutions. To construct the functional network, the brain is parcellated, for example, using  Automated Anatomical Labelling into disjoint anatomical elements (e.g., brain regions) \cite{lee2014hole,tzourio2002automated}. Subsequently, functional information is overlayed on the parcellation to obtain connectivity between the regions. Resting-state functional magnetic resonance imaging (fMRI) is often used to provide a snapshot of the connectivity information over the parcellations at given time points \cite{tzourio2002automated}. Functional connectivity is then computed as the Pearson correlation coefficient between average fMRI time series in the parcellation. The resulting data is commonly presented as a network correlation matrix for further analysis.

The network correlation matrices of the brain networks are often analyzed using conventional graph theory methods \cite{bullmore2009complex}. This involves modeling the brain network as a graph, in which anatomical elements are regarded as nodes, and a connection between any two such regions (nodes) is represented by an edge \cite{bullmore2009complex,honey2007network}. The edges are assigned weights, the correlation measure between the nodes to which it is incident. The topological features of the structural and functional brain networks characterized by graph theory methods are relevant to understanding the brain's various functions and studying brain pathology, both structural and functional. The main interest is often to analyze the most significant connections, which involves thresholding the network matrices to retain only a significant set of connections \cite{dimitriadis2017topological,farahani2019application,wang2010graph,lee2012weighted}. This, however, presents a challenge in brain network analysis when comparing network organization between subjects with a varying number of networks as well as network connectivity strength \cite{wang2010graph,lee2012weighted}. A network template with consistent mathematical properties across networks of different subjects is important for maintaining consistency in comparison and summary measures. The maximum spanning tree (MST) can serve as a reference network measure and increase the comparability between complex networks and various studies.

The MST is a sub-graph that connects all nodes (brain regions) in the network without forming cycles whiles maximizing the edge weights. When the edge weights are unique (which can be assumed in continuous correlation measures), the MST for a given network is unique. Aside from overcoming the arbitrary threshold problem \cite{van2018minimum,dimitriadis2017topological}, the topology of the MST is scale-invariant since its topology depends on the ordering of the weights \cite{tewarie2015minimum}. The robustness of the MST to underlying changes in the original network has made it an effective tool for many network analysis studies. The MST has been used as a reference network to analyze brain disorders such as epilepsy, multiple sclerosis, brain tumors, Alzheimer's disease, and schizophrenia \cite{lee2006classification,smith2015comparison,stam2014trees,tewarie2015minimum,engels2015declining,otte2015aging}. It has also been used to induce over-connectivity to study autism spectrum disorder \cite{benabdallah2020analysis}, and as a reference network to construct graph kernels for measuring the similarity of pairwise brain networks \cite{wang2021multilayer}.

Despite the demonstrated usefulness of MST in many applications, especially in brain network analysis, the MST representation has been limited to the individual subject networks. For example, a group-level MST that captures the topological properties of all the MSTs in a group could provide a more unified and unbiased framework for comparing trees across different network groups. In the literature, a group-level MST is often constructed from the average network or through a bootstrap procedure \cite{ciftci2011minimum,van2018minimum}. The former is biased towards individual subjects with outlying connectivities, and the latter can be computationally prohibitive if a good approximation is to be achieved \cite{de2013estimating,van2018minimum}. To address these issues, we propose a novel spectral representation of trees using the \textit{spanning tree basis} also referred to as the fundamental \textit{cycle basis}. This representation enables us to compute the average MST and demonstrate statistically discriminating localized signals at the edge level and also show that this average tree captures the global properties of all the MSTs in the group.

In summary, to the best of our knowledge, there is no unbiased and computationally efficient method for computing a representation of group-level MSTs. This is the first work to propose a representation of the MST via the spanning tree basis and to explicitly characterize the notion of an average of a collection of trees. The development is theoretically motivated by problems arising in the study of functional brain networks.

\section{Method}
\label{Sec:Methods}

\subsection{Preliminary on trees} 

\subsubsection{Graphs}
\begin{figure}[ht]
	\centering
	\includegraphics[width =\textwidth]{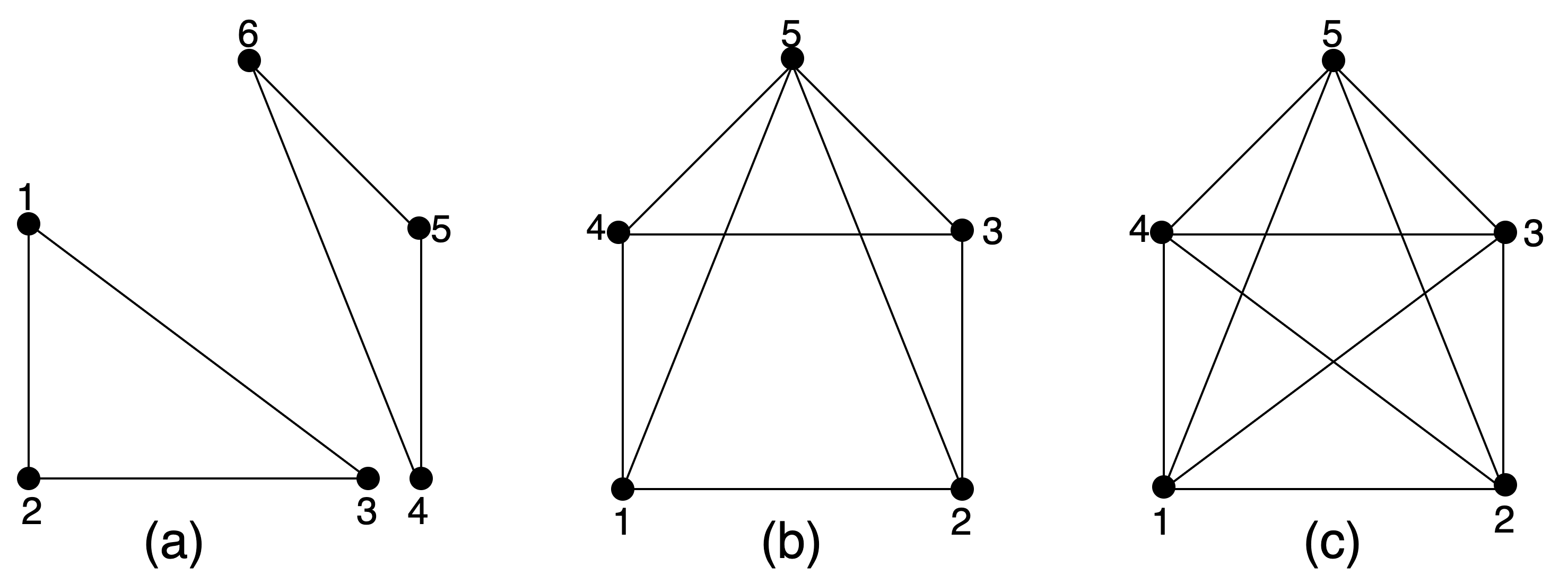}
	\caption{(a) A graph which is not connected. (b) A connected graph. (c) A complete graph. }
	\label{fig:types-of-graphs}
\end{figure}

A graph ${K}$ is an ordered pair ${K} = (\mathcal{V}, \mathcal{E})$ consisting of node set $\mathcal{V}$ and an edge set $\mathcal{E}$. Graphs are often endowed with an incidence function that assigns a weight $\mathit{w}_{i}$ to the edge $(i, j) \in \mathcal{E}$ incident by the nodes $i, j \in \mathcal{V}$. Two nodes that are incident by the same edge are deemed adjacent to each other, and similarly two edges that are incident by a common node are deemed adjacent. Throughout this work, we will denote the number of nodes of a graph by $m$ and the number of edges by $n$. A graph is finite if both $m$ and $n$ are finite.  We will only be concerned with finite graphs which are \textit{connected}. By \textit{connected} we mean a graph which when partitioned into two non-empty sets, there is an edge that connects nodes in the two sets. We also sometimes add the condition that the connected graph be complete, i.e., any two distinct nodes in the graph are adjacent. Figure \ref{fig:types-of-graphs} illustrates the various distinctions.
Since we are concerned with finite graphs, we will define a path in a graph as a finite sequence of edges that join nodes. This leads to the concept of \textit{cycle} (loops) in graphs, which is defined as a non-empty path in which only the first and last vertices are equal. We will assume all the graphs in this work do not contain self-loops, i.e., the first and last vertices in the cycle are distinct. Graphs which do not have cycles are termed \textit{acyclic} graphs \cite{bondy1976graph}. 

\subsubsection{Trees}
In graph theory, trees are defined as \textit{acyclic} graphs. For a tree with $m$ nodes, it has exactly $m-1$ edges, and any two nodes is connected by exactly one edge. A spanning tree is a tree that contains all the nodes in the original graph. Spanning trees are often discussed in the context of connected graphs. A graph that is not connected will not have spanning trees \cite{bondy1976graph}. A graph may have many spanning trees, and this varies based on the type of graph. For a complete graph with $m$ nodes, Cayley's formula \cite{aigner1999proofs} gives the number of spanning trees as $m^{m-2}$. A maximum spanning tree (MST) of a weighted connected graph is a spanning tree that maximizes the total edge weights of the tree. The Kruskals algorithm, which was designed primarily to construct minimum spanning trees can be adapted to construct MSTs \cite{kruskal1956shortest}.

\subsection{Hodge Laplacian over graphs}

\subsubsection{Boundary matrix of a graph} 
The graph $K$ is sometimes referred to as the 1-skeleton, consisting of nodes ($0$-simplices) and edges ($1$-simplices). A $1$-simplex $e_{ij} = (i, j)$, is a $1$-dimensional convex hull (polytope) of $2$ points $v_i, v_j$. For simplices $\{\sigma_i \in K\}$, the finite sum $\sum_i a_i \sigma_i$, is referred to as a $0$-chain if $\sigma_i$ are $0$-simplices and a $1$-chain if $\sigma_i$ are $1$-simplices. Here the coefficients $a_i$ are either $0$ or $1$. The set of $0$-chains forms a group and a sequence of these groups is called a chain complex denoted $C_0$. Similarly, the set of $1$-chains forms a group and is denoted as $C_1$. To relate the $1$-chain group to the $0$-chain group, we define the boundary map $\partial: C_1 \longrightarrow C_{0}$ as follows \cite{topaz2015topological}

\begin{equation}
    \partial(e_{ij}) = v_j - v_i, \quad i < j.
    \label{eqn:boundary-operator}
\end{equation}
Let $(\mathbb{B}_{k, (i,j)})$ denote a matrix whose rows are indexed by nodes $k$ and the columns indexed by edges $e_{ij} = (i, j)$. The matrix representation $\mathbb{B} = (\mathbb{B}_{k, {(i,j)}})$ of the boundary operator $\partial$, is given as 
\begin{equation}
	\mathbb{B}_{k, {(i,j)}} =
	\begin{cases}
		1, 	& \text{if } v_k = v_i\\
		-1, & \text{if } v_k = v_j\\
		0,  & \text{otherwise}.
	\end{cases},
	\label{eqn:boundary-matrix}
\end{equation}
The boundary operator generalizes the incidence matrix in graph theory \cite{kaczynski2004computational}.

\subsubsection{Hodge Laplacian} 

The Hodge Laplacian matrix $\mathcal{L}$ of $K$ is defined using the boundary matrix, which is the matrix form of the boundary operator: 
\begin{equation}
	\mathcal{L} =	\mathbb{B}^T\mathbb{B}. 
	\label{eqn:HodgeLaplacian}	
\end{equation}
Note that the Hodge Laplacian is a higher dimensional generalization of the combinatorial graph Laplacian. In spectral graph theory, the combinatorial graph Laplacian and its eigendecomposition has been used to characterize various graph properties. The eigenvector associated with the zero eigenvalue has been useful in clustering task \cite{von2007tutorial}. More importantly and quite pertinent to this work, the  eigenvectors of the the graph Laplacian can serve as a useful basis for representing a graph when data is defined on the graph nodes \cite{ortega2018graph,bach2013learning,jost2021discrete,barbarossa2020topological,dakurah2025discrete}. We extend this idea to the Hodge Laplacian and propose a basis representation for trees using the eigenvectors of the Hodge Laplacian when the data is defined on tree edges.

\subsection{Algebraic representation of cycles and trees}

Comparing trees across different networks requires a standardized representation that is consistent across different networks. Just as in many fields of algebra, a basis can serve as the standardized representation. In what follows, we show how to construct the spanning tree basis as a way of establishing a unified framework for tree comparisons.

\subsubsection{Fundamental cycle basis as spanning tree basis}
We can construct the \textit{spanning tree basis} in the graph $K$ as follows. We assume $K$ is a connected complete graph. The edge set of $K$ is denoted as $\mathcal{E}(K)$. For every spanning tree $T$ of $K$, there is a basis $\mathcal{C}$, 
which is the set of \textit{cycles} formed by adding an edge $e_{ij} = (i, j) \in \mathcal{E}(K')$ to $T$, where
$K'$ denotes a subgraph that is not part of the spanning tree.
$\mathcal{C}$ is the basis over the collection of every possible cycles \cite{dakurah2022modelling,dakurah2025robust}.
As each pair of distinct vertices in $K$ are joined by unique path in $T$, the subgraph $T \cup e_{ij}$ has a unique cycle. Each of these cycles has an edge contained by no other cycle, so the set of cycles must be independent. The spanning tree $T$ has $m-1$ edges and the cardinality of $\mathcal{C}$ is $n-m+1$, hence $\mathcal{C}$ is maximally independent such that the cardinality of $\mathcal{C}$ can not be increased without inducing dependence. This basis is referred to as the \textit{spanning tree basis} \cite{barr1979enhancements,zach2010disambiguating,dancso2016construction}.

\begin{figure}[ht]
	\centering
	\includegraphics[width =\textwidth]{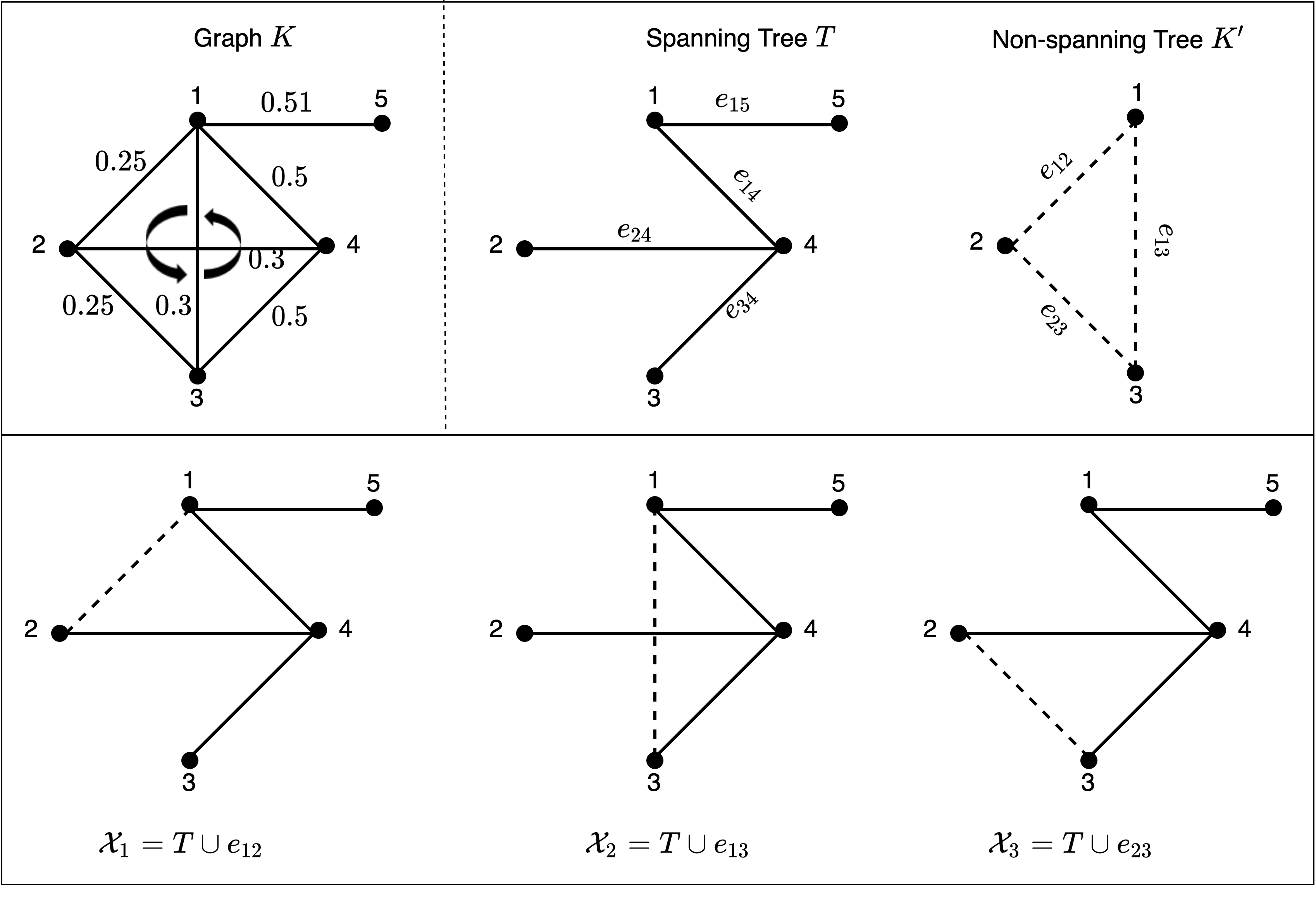}
	\caption{Top: A connected graph $K$ decomposed into a spanning tree $T$ and a non-spanning tree $K^\prime$. Bottom: Subgraphs constructed by adding edges from $K^\prime$ to $T$. The cycles in the subgraphs will be identified using the Hodge Laplacian.}
	\label{fig:unique-subgraphs}
\end{figure}

The Hodge Laplacian can be used to build the set $\mathcal{C}$. The elements of $\mathcal{C}$ are contained in the $1$-st homology group $\mathcal{H} = ker\mathcal{L}$, i.e., the kernel of Hodge Laplacian $\mathcal{L}$ \cite{lee2014hole}. The eigenvectors with zero eigenvalues of $\mathcal{L}$ span the kernel space of $\mathcal{L}$. The eigendecomposition of $\mathcal{L}$ is given by 
\begin{equation} 
\mathcal{L} {\Psi} = {\Psi} {\Lambda},
\label{eqn:eig-decomposition}
\end{equation}
where ${\Lambda} = diag(\lambda_1, \cdots, \lambda_n)$ is a diagonal matrix with the eigenvalues of $\mathcal{L}$ as diagonal entries and  corresponding eigenvectors as the columns of the matrix ${\Psi} = [\psi_1, \cdots, \psi_n]$. The cardinality of $\mathcal{C}$ is given by the multiplicity of the zero eigenvalues, and is also referred to as the $1$-st Betti number $\beta_1 = n-m+1$.

To construct $\mathcal{C}$, we proceed sequentially as follows.  A sequence of unique subgraphs $\mathcal{X}_k = \{T \cup e_{ij}: e_{ij}\in \mathcal{E}(K^\prime)\}$ are constructed. See Figure \ref{fig:unique-subgraphs} for this subgraphs construction process. The Hodge Laplacian for each unique subgraph will yield an eigendecomposition where the multiplicity of the zero eigenvalue will be exactly one. Further, the entries of the corresponding eigenvector will have non-zero values only for those edges that constitute the cycle and the rest of the entries are zeros. Without loss of generality, we order the eigenvalues $0=\lambda_1 < \lambda_2 \le  \cdots \le \lambda_{n}$. Then the eigenvector that identifies the cycle in $\mathcal{X}_k$ is given by $\psi_1$. Observe that the entries of the eigenvectors are indexed by $(i,j)$, in a similar fashion as the columns of the boundary matrix in \ref{eqn:boundary-matrix}. For example, for the first subgraph $\mathcal{X}_1$ in Figure \ref{fig:unique-subgraphs}, $\psi_1$ will have the following representation (see Example \ref{exp:example1} for a complete illustration):
$$
\psi_1 = 
\begin{blockarray}{ccccccc}
    & & e_{12} & e_{14} & e_{15} & e_{24} & e_{34} \\
    \begin{block}{cc[ccccc]}
        & & \psi_{1, ({1,2})} & \psi_{1, ({1,4})} & \psi_{1, ({1,5})} & \psi_{1, ({2,4})} & \psi_{1, ({3,4})}\\
    \end{block}
\end{blockarray}
$$
Note that this representation only has the edges in the subgraph. To properly represent the basis, we augment the above to include all the edges in the graph with a slight abuse of notation were we denote the augmented $\psi_1$ with $\psi_1$:
$$
\psi_1 = 
\begin{blockarray}{ccccccccc}
    & & e_{12} & e_{13} & e_{14} & e_{15} & e_{23} & e_{24} & e_{34} \\
    \begin{block}{cc[ccccccc]}
        & & \psi_{1, ({1,2})} & 0 & \psi_{1, ({1,4})} & \psi_{1, ({1,5})} & 0 & \psi_{1, ({2,4})} & \psi_{1, ({3,4})}\\
    \end{block}
\end{blockarray}
$$ 
Then the cycle in the graph  $\mathcal{X}_k$ is algebraically represented as

\begin{equation}    
\mathcal{C}_k = \sum_{e_{ij} \in \mathcal{E}(K)} e_{ij} \psi_{1, (i,j)}.
\label{eqn:k-cycles-relevant} 
\end{equation}

A full demonstration of this representation is provided in Example \ref{exp:example1}. Now, for all such subgraphs, the set $\mathcal{C}$ is given by $\mathcal{C} = \{ \mathcal{C}_1, \cdots, \mathcal{C}_{\beta_1} \}$. The set $\mathcal{C}$ form the \textit{spanning tree basis}.

\begin{example}
Consider the graph $K$ and the corresponding spanning and non-spanning tree in Figure \ref{fig:unique-subgraphs}. Take the first subgraph $\mathcal{X}_1 = T \cup e_{12}$, its boundary matrix and the associated Hodge Laplacian are given by

$$
\mathbb{B} = 
\arraycolsep=54.5pt
\begin{blockarray}{ccccccc}
    & & e_{12} & e_{14} & e_{15} & e_{24} & e_{34}\\
    \begin{block}{cc(ccccc)}
        \left[1\right] & & 1 & 1 & \sminus 1 & 0 & 0\\
         \left[2\right] & & \sminus 1 & 0 & 0 & 1 & 0\\
        \left[3\right] & & 0 & 0 & 0 & 0 & 1\\
        \left[4\right] & & 0 & \sminus 1 & 0 & \sminus 1 & \sminus 1\\
        \left[5\right] & & 0 & 0 & \sminus 1 & 0 & 0\\
    \end{block}
\end{blockarray}
\quad, \quad
\setlength\arraycolsep{6pt}
\mathcal{L} = \mathbb{B}^T\mathbb{B}=
\begin{pmatrix}
        2 & 1 & 1 & - 1 & 0\\
        1 & 2 & 1 & 1 & 1\\
        1 & 1 & 2 & 0 & 0\\
        -  1 & 1 & 0 & 2 & 1\\
        0 & 1 & 0 & 1 & 2 \\
\end{pmatrix}.
$$

The eigendecomposition $\mathcal{L} \Psi= \Psi \Lambda$ will yield
$$
\arraycolsep=3.7pt\def\arraystretch{1}
\Psi = 
\begin{blockarray}{ccccccc}
    & & \psi_1 & \psi_2 & \psi_3 & \psi_4 & \psi_5\\
    \begin{block}{cc(ccccc)}
        e_{12} & & \mathbf{0.58} & \sminus 0.25 & 0.37 & \sminus 0.60 & 0.33\\
        e_{14} & & \mathbf{\sminus 0.58} & \sminus 0.49 & 0.00 & 0.00 & 0.65\\
        e_{15} & & \mathbf{0.00} & 0.57 & \sminus 0.60 & \sminus 0.37 & 0.42\\
        e_{24} & & \mathbf{0.58} & \sminus 0.25 & \sminus 0.37 & 0.60 & 0.22\\
        e_{34} & & \mathbf{0.00} & 0.57 & 0.60 & 0.37 & 0.43\\
    \end{block}
\end{blockarray}
\quad, \quad
\arraycolsep=1.0pt\def\arraystretch{1}
\Lambda = 
\begin{pmatrix}
        \mathbf{0.00} & 0 & 0 & 0 & 0\\
        0 & 0.70 & 0 & 0 & 0\\
        0 & 0 & 1.38 & 0 & 0\\
        0 & 0 & 0 & 3.62 & 0\\
        0 & 0 & 0 & 0 & 4.30\\
\end{pmatrix},
$$
The eigenvector $\psi_1$ corresponds to the zero eigenvalue. The cycle in $\mathcal{X}_1$ can then be represented as 
$$
\mathcal{C}_1 = 0.58 e_{12} + 0.00 e_{13}  - 0.58 e_{14} + 0.00 e_{15} + 0.00e_{23} + 0.58 e_{24} + 0.00 e_{34}
$$
Similarly, the cycle in the subgraphs $\mathcal{X}_2 = T \cup e_{13}$ and $\mathcal{X}_3 = T \cup e_{23}$ is identified and represented respectively as
$$
\mathcal{C}_2 = 0.00 e_{12} - 0.58 e_{13}  + 0.58 e_{14} + 0.00 e_{15} + 0.00e_{23} + 000 e_{24} - 0.58 e_{34}
$$
$$
\mathcal{C}_3 = 0.00 e_{12} + 0.00 e_{13}  + 0.00 e_{14} + 0.00 e_{15} + 0.58e_{23} - 0.58 e_{24} + 0.58 e_{34}
$$
The coefficients of these cycles will form the sparse matrix $\Phi = \left[\mathcal{C}_1, \mathcal{C}_2, \mathcal{C}_3\right]$ with the form

$$
\Phi = 
\begin{pmatrix}
    0.58 & 0.00 & 0.00 \\
    0.00 & \sminus 0.58 & 0.00 \\
    \sminus 0.58 & 0.58 & 0.00 \\
    0.00 & 0.00 & 0.00 \\
    0.00 & 0.00 & 0.58 \\
    0.58 & 0.00 & \sminus 0.58 \\
    0.00 & \sminus 0.58 & 0.58 \\
\end{pmatrix}
\quad
\xrightarrow[]{\text{equivalent basis}}
\quad
\begin{pmatrix}
    1.00 & 0.00 & 0.00 \\
    0.00 & 1.00 & 0.00 \\
    1.00 & 1.00 & 0.00 \\
    0.00 & 0.00 & 0.00 \\
    0.00 & 0.00 & 1.00 \\
    1.00 & 0.00 & 1.00 \\
    0.00 & 1.00 & 1.00 \\
\end{pmatrix}
$$
This standardized representation can be used to analyze various features across multiple spanning trees. For convenience and in alignment with our theoretical development, we will use the equivalent basis representation for the rest of this paper.
\label{exp:example1}
\end{example}

\subsection{The average spectral tree} 
Trees do not have a natural representation in the vector space. This makes summary statistical measures such as mean and variance difficult to define and characterize. For a collection of edges for a given tree, it is possible to define statistical quantities on the edges. This however can not be directly extended to trees since it is not clear how the edges of a group of trees are related. The spectral decomposition of a tree can serves as a useful tool for relating different edges of graphs. Through the spectral representation, we can develop an averaging process which can be use to compute a tree which is as close as possible to the average tree.

\subsubsection{Spectral representation of trees}
The spectral representation of a tree can be obtained by writing it as a combination of the spanning tree basis obtained in the previous section. Given a collection of networks $\mathbb{K} = \{K_1, \cdots, K_N\}$, assume these are complete graphs. Denote their corresponding spanning trees by $\mathbb{T} = \{ T_1, \cdots, T_N \}$. Each spanning tree will have a lexicographical (based on the node labels) vectorized representation where the entries will be the edge weights of it's connectivity matrix. We will assume the edge weights are all positive. Note that in this representation, edges not in the particular tree will have their entries been zero. To construct the spectral representation of the trees, we proceed as follows. 

The average network can be obtained by averaging the connectivity matrices of all the networks in $\mathbb{K}$. Let $\Bar{K}$ denote this average network. We can obtain the fundamental spanning tree basis associated with this network according to \ref{eqn:k-cycles-relevant}, denoted as $\Phi = \left[ \mathcal{C}_1, \cdots, \mathcal{C}_{\beta_1} \right]$. Then for any tree $T_k \in \mathbb{T}$, we can represent it in terms of the basis via the minimization problem

\begin{equation}
     \argmin_{{\alpha}^k \ge 0} \left( \frac{1}{2} (\alpha^k)^\top Q \alpha^k + u^\top \alpha^k \right),
     \label{eqn:minimization-prob}
\end{equation}

where ${\alpha^k} = \left[\alpha_1^k, \cdots, \alpha_{\beta_1}^k \right]^{\top}$ are the  spectral expansion coefficients, $Q = \Phi^\top \Phi$ and $u = -\Phi^\top T_k$. This is a convex optimization problem and the non-negative constraints is in keeping with the algebraic structure of $T_k$. Then the tree $T_k$ admits an algebraic expansion into the basis as follows
\begin{equation}
    T_k \approx \sum_{l=1}^{\beta_1} \hat{\alpha}_l^k \mathcal{C}_l.
    \label{eqn:alg-trees}
\end{equation}
Note that we are assuming $T_k$ to be it's algebraic representation. We demonstrate this representation process in the following example.

\begin{example}
Consider the three networks given in Figure \ref{fig:different-tree-topology}. Their corresponding maximum spanning trees are below each network.
\begin{figure}[ht]
	\centering
	\includegraphics[width =\textwidth]{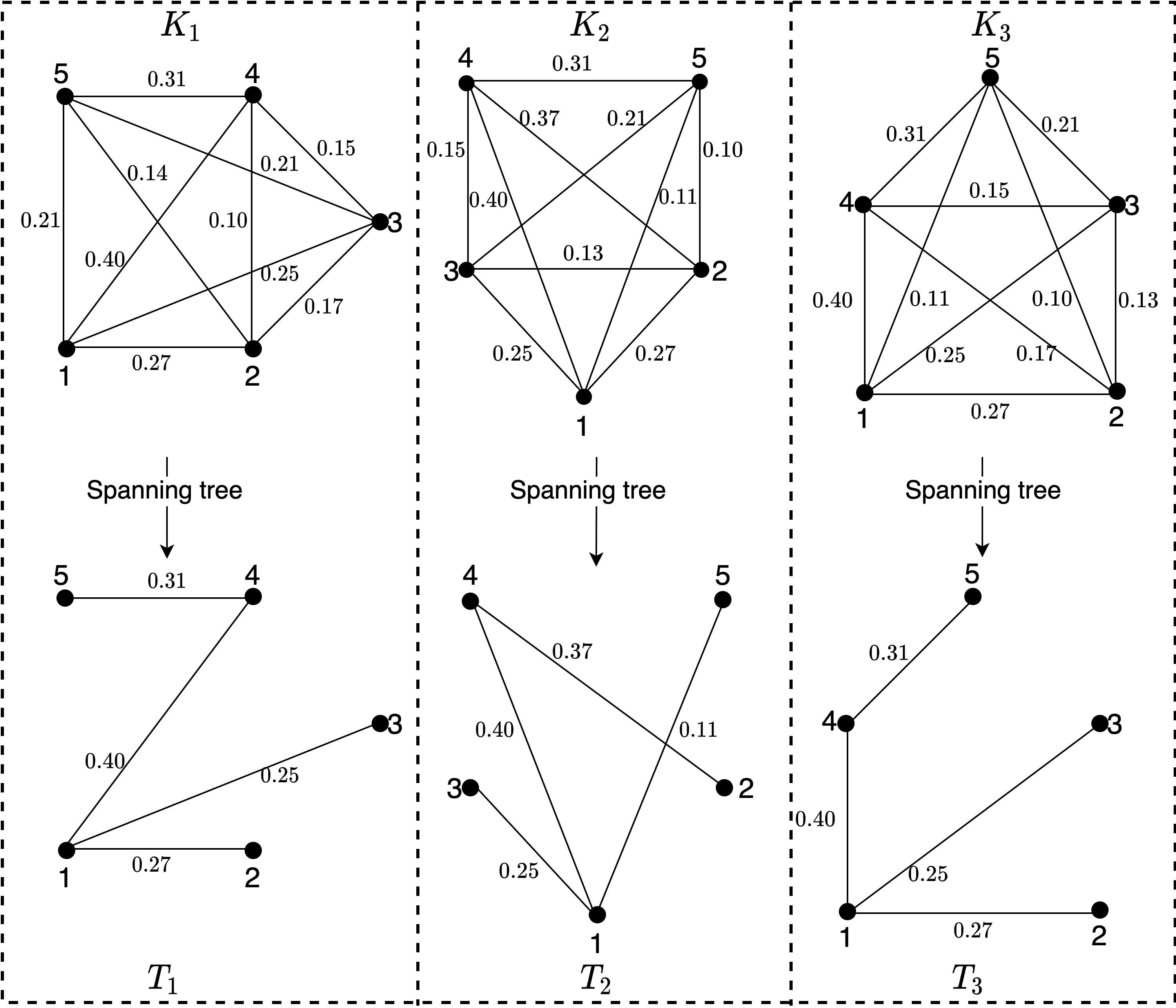}
	\caption{Three networks with their corresponding spanning maximum spanning trees.}
	\label{fig:different-tree-topology}
\end{figure}
To compute the basis and derive the algebraic representation of the trees we first compute the average connectivity matrices of the three networks as illustrated in Figure \ref{fig:average-network}.
\begin{figure}[ht]
	\centering
	\includegraphics[width =\textwidth]{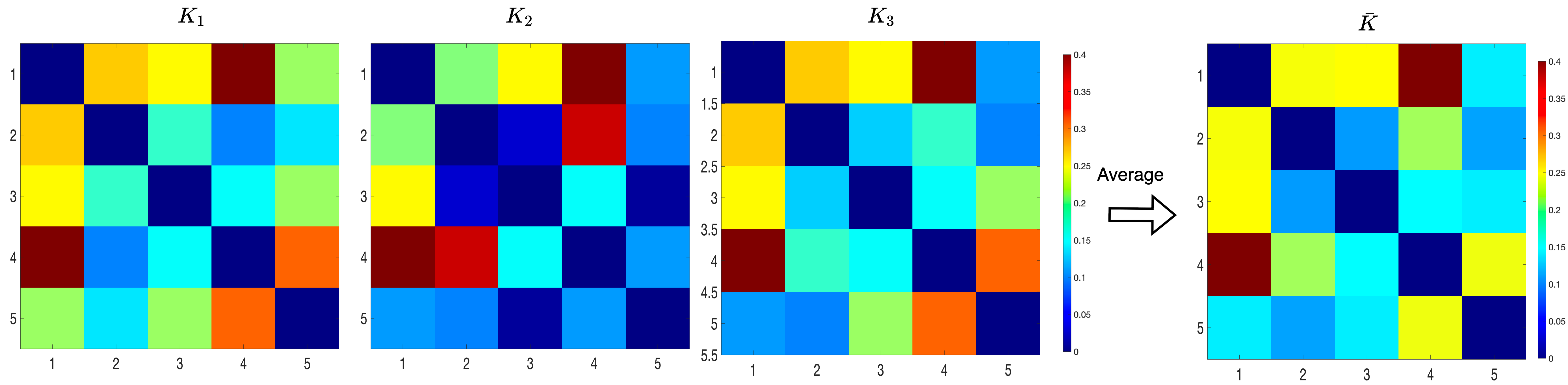}
	\caption{The connectivities of the three networks and their average connectivity.}
	\label{fig:average-network}
\end{figure}
The lexicographical vectorized representation of the trees and the corresponding spanning tree basis from the average network is given below.
$$
T_1 = 
\begin{blockarray}{c}
\begin{block}{(c)}
    0.27\\
    0.25\\
    0.40\\
    0.00\\
    0.00\\
    0.00\\
    0.00\\
    0.00\\
    0.00\\
    0.31\\
\end{block}
\end{blockarray}
\quad,\quad
T_2 = 
\begin{blockarray}{c}
\begin{block}{(c)}
    0.00\\
    0.25\\
    0.40\\
    0.11\\
    0.00\\
    0.37\\
    0.00\\
    0.00\\
    0.00\\
    0.00\\
\end{block}
\end{blockarray}
\quad,\quad
T_3 = 
\begin{blockarray}{c}
\begin{block}{(c)}
    0.27\\
    0.25\\
    0.40\\
    0.00\\
    0.00\\
    0.00\\
    0.00\\
    0.00\\
    0.00\\
    0.31\\
\end{block}
\end{blockarray}
\quad,\quad
\Phi = 
\begin{blockarray}{c c c c c c}
\begin{block}{(c c c c c c)}
    0.00 & 1.00 & 1.00 & 1.00 & 0.00 & 0.00 \\
    0.00 & 1.00 & 0.00 & 0.00 & 1.00 & 1.00 \\
    1.00 & 0.00 & 1.00 & 1.00 & 1.00 & 1.00 \\
    1.00 & 0.00 & 0.00 & 0.00 & 0.00 & 0.00 \\
    0.00 & 1.00 & 0.00 & 0.00 & 0.00 & 0.00 \\
    0.00 & 0.00 & 1.00 & 0.00 & 0.00 & 0.00 \\
    0.00 & 0.00 & 0.00 & 1.00 & 0.00 & 0.00 \\
    0.00 & 0.00 & 0.00 & 0.00 & 1.00 & 0.00 \\
    0.00 & 0.00 & 0.00 & 0.00 & 0.00 & 1.00 \\
    1.00 & 0.00 & 0.00 & 1.00 & 0.00 & 1.00 \\
\end{block}
\end{blockarray}
\quad .
$$
Then using the representation in \ref{eqn:minimization-prob}, we can estimate the coefficients associated with each of the three trees as 
$$
\hat{\alpha}^1 = 
\begin{blockarray}{c}
\begin{block}{(c)}
    0.07\\
    0.07\\
    0.06\\
    0.10\\
    0.05\\
    0.10\\
\end{block}
\end{blockarray}
\quad,\quad
\hat{\alpha}^2 = 
\begin{blockarray}{c}
\begin{block}{(c)}
    0.04\\
    0.00\\
    0.19\\
    0.00\\
    0.11\\
    0.04\\
\end{block}
\end{blockarray}
\quad,\quad
\hat{\alpha}^3 = 
\begin{blockarray}{c}
\begin{block}{(c)}
    0.07\\
    0.07\\
    0.06\\
    0.10\\
    0.05\\
    0.10\\
\end{block}
\end{blockarray}
\quad .
$$
Here, $\hat{\alpha}^1$, $\hat{\alpha}^2$ and, $\hat{\alpha}^3$ are the estimated spectral expansion coefficients corresponding to the trees $T_1$, $T_2$ and $T_3$ respectively. Using these expansion coefficients, we can develop an averaging process.
\label{exp:average-network-basis}
\end{example}

\subsubsection{Computing the average spectral tree} We propose the following procedure for obtaining the spectral average of a collection of trees. Given the collection of trees $\mathbb{T} = \{T_1, \cdots, T_N\}$, we compute the average of their algebraic representation given in \ref{eqn:alg-trees} as 
\begin{equation}
    \frac{1}{N} \left( \sum_{l=1}^{\beta_1} \hat{\alpha}_l^1 \mathcal{C}_l + \cdots + \sum_{l=1}^{\beta_1} \hat{\alpha}_l^N \mathcal{C}_l \right).
\end{equation}
We can compute the component-wise mean spectral expansion coefficients as follows

\begin{equation}
    \bar{\alpha}_l =  \frac{1}{N} \sum_{i = 1}^N \hat{\alpha}_l^i, \quad \quad l = 1, \cdots, \beta_1.
    \label{eqn:avg-coefficients}
\end{equation}
Let $\bar{T}$ denote the average tree. We propose the following algebraic representation of $\bar{T}$:
\begin{equation}
    \bar{T} = \sum_{l=1}^{\beta_1} \bar{\alpha}_l \mathcal{C}_l.
    \label{eqn:avg-tree}
\end{equation}
The resulting $\bar{T}$ might not correspond exactly to a spanning tree and it will be necessary to project it to the nearest tree \cite{demetci2020gromov,maretic2020wasserstein,umeyama1988eigendecomposition,white2006mixing}.

\begin{example}
As a continuation of Example \ref{exp:average-network-basis}, we have already computed the spectral expansion coefficients for each tree in Figure \ref{fig:different-tree-topology}. 
Using the average tree formula in equation \ref{eqn:avg-tree}, we can now compute the average of the three trees as illustrated in Figure \ref{fig:average-tree}.

\begin{figure}[ht]
	\centering
	\includegraphics[width =\textwidth]{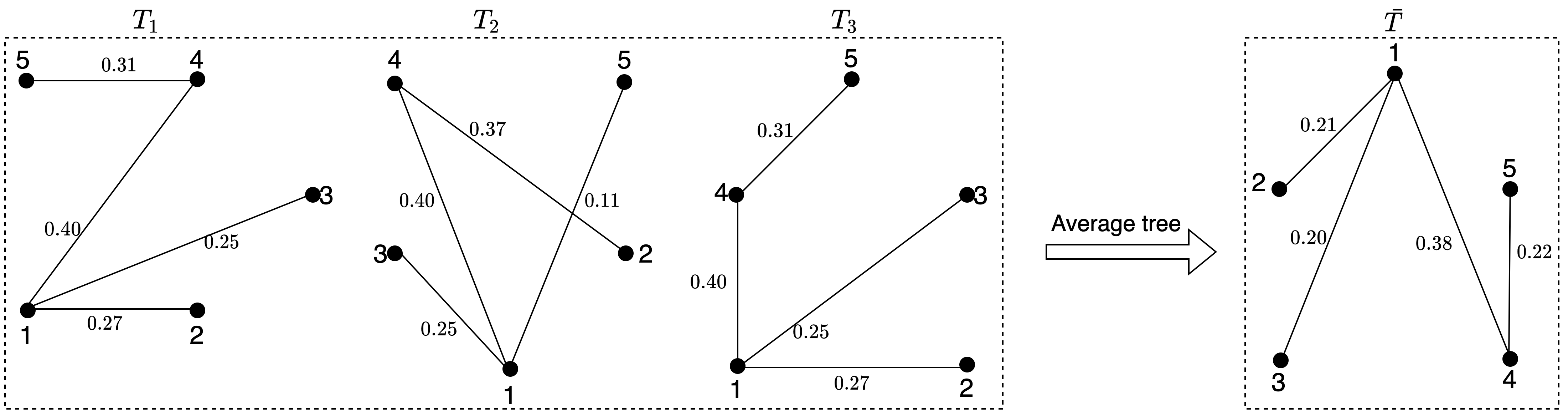}
	\caption{Illustration of the average tree computation. On the left are the tree spanning trees constructed from the network. The right denotes the computed average tree. Note that the geometry of the tree is not unique and the projection was unto a predefined template that matches the geometry of the original set of trees.}
	\label{fig:average-tree}
\end{figure}
\label{exp:average-tree-computation}
\end{example}

\subsection{Shortest-path trees}

To identify and analyze the frequently used shortest communication paths for information transfer in the networks, we introduce the concept of shortest-path trees.

\subsubsection{Shortest paths in networks} In previous sections, we defined a path in a network as as a finite sequence of edges that join nodes. More formally, assume the network $K$ is endowed with a weight function $w_{ij}$ for $i, j \in \mathcal{V}$. Let $s$ be a source node and $t$ a target node. The path $\Omega_{s \xrightarrow[]{} t}$ is defined as the sequence of distinct nodes joining the source node ($s$) and the target node ($t$):

\begin{equation}
    \Omega_{s \xrightarrow[]{} t} = \{ s, i, j, \cdots, k, t \}.
\end{equation}
In functional brain networks, edges with large weights are the strongest and most reliable connection among the brain regions \cite{park2013structural}. Computing the shortest path, to be defined shortly, on the functional network will include less influential edges. A monotonic transformation of the weight of the edges is necessary before computing the shortest path. To this end, define $d_{ij} = 1/w_{ij}$ to be the distance between node $i$ and node $j$. The shortest-path $\pi_{s \xrightarrow{} t}$ from a source node ($s$) to a target node ($t$) is then defined as a path that minimizes the distance between the two nodes. This path is comprised of the sequence 
\begin{equation}
    \pi_{s \xrightarrow{} t} = \{d_{si}, d_{ij}, \cdots, d_{kt}\}.
\end{equation}
The Dijstra's algorithm is often used to compute the shortest path \cite{mieghem2006data}.
\begin{figure}[ht]
	\centering
	\includegraphics[width =\textwidth]{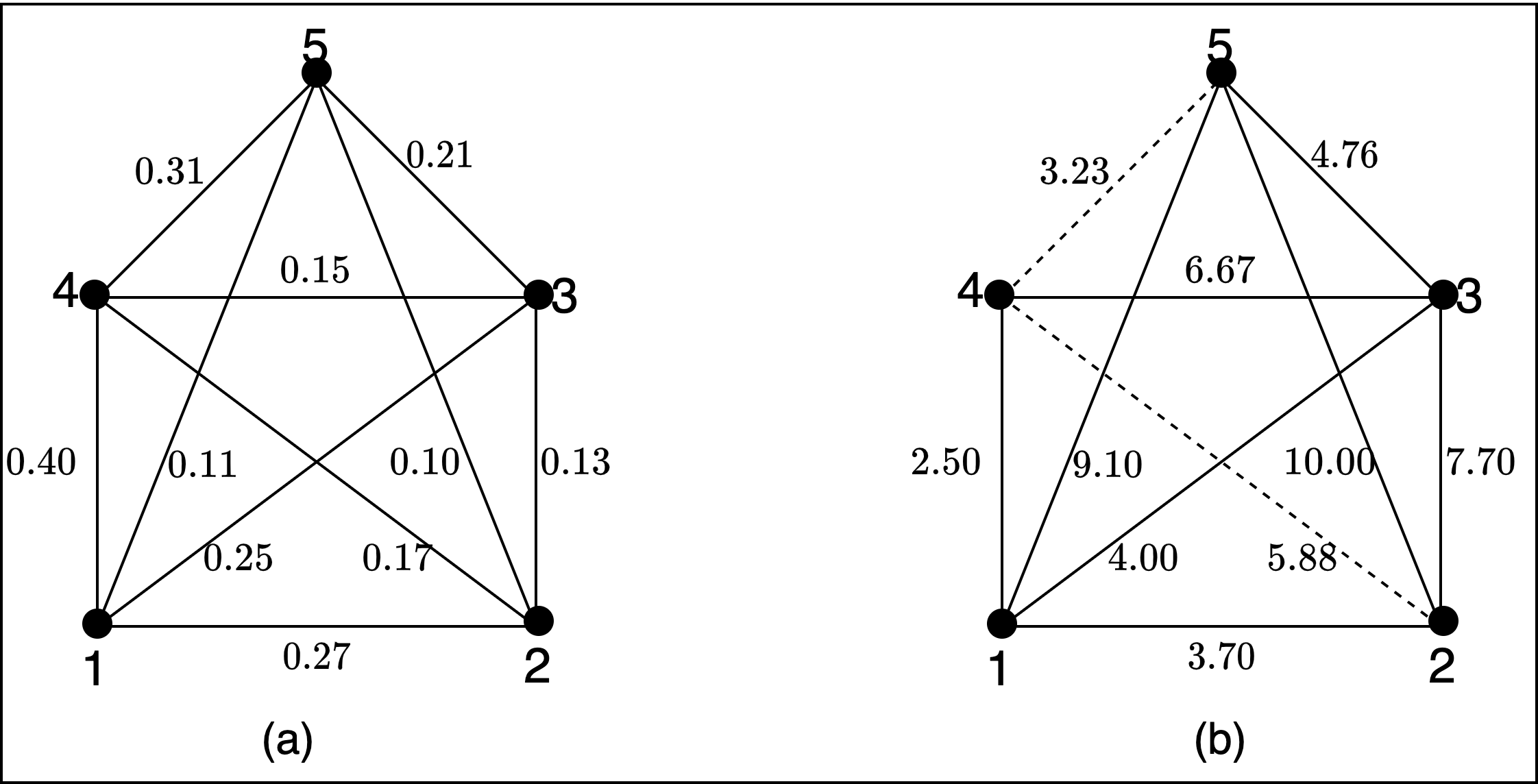}
	\caption{Illustration of the shortest path computation. (a) A complete weighted network. (b) A complete network with the edge weights been the measure $d_{ij}$. The shortest-path between node 2 and 5 is denoted by the dashed line.}
	\label{fig:shortest-path-illustration}
\end{figure}
Figure \ref{fig:shortest-path-illustration} illustrates an example of the shortest-path computation in a complete network. The shortest-path between the source node $2$ and the target node $5$ denoted $\pi_{2 \xrightarrow{} 5}$ is given by
$$
\pi_{2 \xrightarrow{} 5} = \{d_{24}, d_{45}\}
$$
We can represent the node sequence associated with this path as 
$\Omega_{2 \xrightarrow{} 5} = \{2, 4, 5\}$. In general, there can be more than one shortest-path between any pair of nodes \cite{mieghem2006data,van2008observable}.

\subsubsection{Shortest-path tree}
Computing all the shortest-path of a complete network can be intuitively thought of as extracting the backbone or the main highway (communication paths) of the network. For a given node, the union of all the shortest paths from that node to all the other nodes in the network will form a tree \cite{van2010graph}. The shortest-path tree rooted at a given node is therefore defined as the union of the shortest paths from that particular node to all other nodes in the network\cite{van2010graph,meier2015union}.
\begin{figure}[ht]
	\centering
	\includegraphics[width =\textwidth]{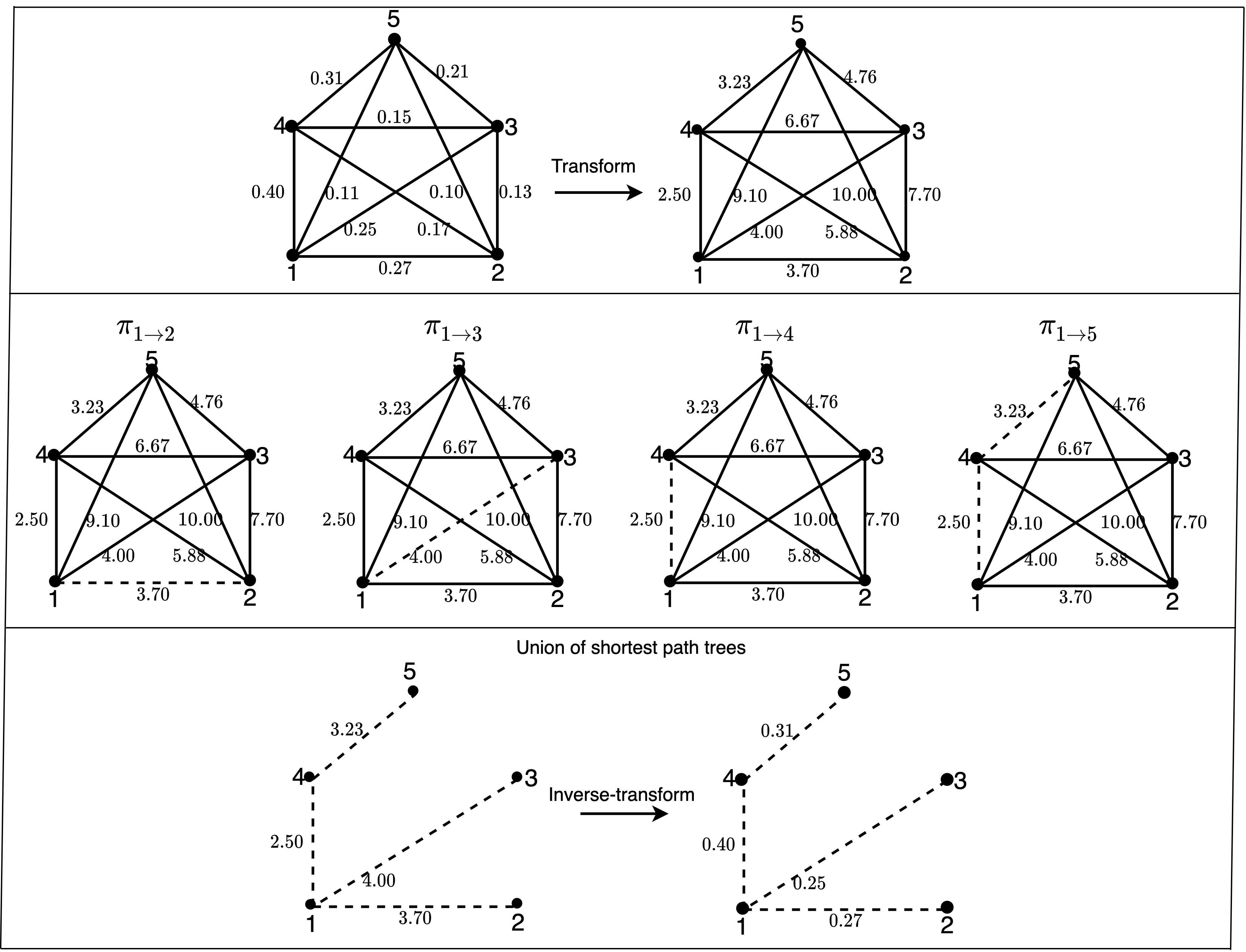}
	\caption{Illustration of the shortest path tree computation. Top: a complete network and its weight transform. Middle: The shortest paths from node 1 to all other nodes. Bottom: The union of the shortest paths and it's inverse transform.}
	\label{fig:shortest-path-tree}
\end{figure}
The shortest-path tree computation is illustrated in Figure \ref{fig:shortest-path-tree}. In this particular case, we are computing the shortest-path tree rooted at node 1. To do so, we first compute the shortest-path from node 1 to each of the other nodes in the network. We then take the union of all the paths. This results in the shortest-path tree. Coincidentally, this shortest-path tree is the maximum spanning tree. This is a very important realization and we will formalize it in the next section. 

\subsubsection{The union of shortest-path trees}
For a given network $K$, we can construct the shortest-path trees associated with all the nodes in the network. If we take the union of all such shortest path trees, we will have a subgraph $\mathcal{U} \subseteq K$. The subgraph $\mathcal{U}$ is the transport hub with regards to the topology of $K$, that is $\mathcal{U}$ is the maximally observable part of $K$ through which through which most information flow in the network revolves around \cite{van2008observable,meier2015union}. Observe that if all the edge weights of the network $K$ are unitary, then the subgraph $\mathcal{U}$ is exactly equal to the graph $K$.  We state an important result that tie together the relationship between the network, the union of the shortest path trees and the maximum spanning tree \cite{van2008observable}.

\begin{theorem}
Let $K$ be a complete weighted network and $\mathcal{U}$ the union of the shortest path trees associated with the nodes of $K$. Also let $T$ be the maximum spanning tree associated with the network $K$. Then the following holds:
$$K \supseteq \mathcal{U} \supseteq T.$$
\end{theorem}

This theorem essentially shows that the maximum spanning tree is a subgraph of the union of the shortest-path trees. In the special case where $\mathcal{U} = T$, all the traffic in the network flows through the edges of $T$. But when $\mathcal{U} \supset T$, information flows through other edges that are not a part of $T$.

\subsection{Efficiency of the shortest paths}
The underlying principle of the shortest paths is the efficiency with which they exchange information in the network \cite{latora2001efficient}. We can quantify this efficiency by defining and characterizing the notion of path information \cite{sneppen2005hide,trusina2005communication}. Essentially path information measures the accessibility of the path linking a source node to a target node. This measure depends on the node degree along the path \cite{sneppen2005hide,trusina2005communication}. Here, our focus will be the information along the shortest paths. Networks with smaller path information have more concentrated flow of information along the major highways compared to networks with large path information.
For each node $i$ in the network, we define the node strength as follows:
\begin{equation}
    s_i = \sum_{i \ne j} w_{ij}.
\end{equation}
This is the weighted node degree \cite{bondy1976graph}. If we let $\pi_{i \xrightarrow{} t}^{(1)}$ represent the first element of the path $\pi_{i \xrightarrow{} t}$ and $\Omega^*_{s\xrightarrow{}t}$ represent the sequence of nodes excluding the target, that is $\Omega^*_{s\xrightarrow{}t} = \{s, i, j, \cdots, k\}$. Then the probability of taking the shortest path from $s$ to $t$ may be expressed as
\begin{equation}
    P(\pi_{s \xrightarrow{} t}) = \prod_{i \in \Omega^*_{s \xrightarrow{} t}} \left( \frac{1}{\pi^{(1)}_{i \xrightarrow{} t} \times s_i} \right).
\end{equation}
Since we are dealing with fully connected networks, we assume the shortest path between any two nodes always exists. Then the information needed to access the target node $t$ from the source node $s$ is given by

\begin{equation}
    I(s, t) = -\log(P(\pi_{s \xrightarrow{} t})).
    \label{eqn:infor-shortest-path}
\end{equation}
First we note that the strength of a node will affect the efficiency of information flow, which is largely influenced by the node degree. What this measure $I(, t)$ quantifies is that paths with stronger edge weights, involving fewer links with lower node degree would promote efficient information flow. 

\section{Statistical inference}
The pertinent statistical issue we wish to address is to summarize and compare group of trees in a a statistically coherent manner\cite{dakurah2025topologically,dakurah2025maxtda}. We will also develop an  inference procedure for detecting difference in the organization of information flow between networks based on the path information introduced in the previous section.

\subsection{Individual tree comparison}
Given a collection of networks in two groups, we can compute the MST associated with each network in either group. Let $\mathbb{T}_1$ and $\mathbb{T}_2$ be the collection of trees in the two groups. We can now test whether the collection of trees in the two groups differ. The test procedure will be developed around the spectral coefficients of each tree given in \ref{eqn:alg-trees}.
The statistical inference is performed on the coefficients. Let $\bar{\alpha}^1_l$ and $\bar{\alpha}^2_l$ be the mean coefficients corresponding to the $l$-$th$ \textit{spanning tree basis} of networks in group $\mathbb{T}_1$ and group $\mathbb{T}_2$ respectively. 
We propose the following statistic for testing the difference in the distribution of the trees between the two groups:
\begin{equation}
    \mathcal{T}(\mathbb{T}_1, \mathbb{T}_2) = \max_{1 \le l \le \beta_1} |\bar{\alpha}^1_l - \bar{\alpha}^2_l|
    \label{eqn:cycle-coefficients}
\end{equation}
The statistical significance is determined using the permutation test \cite{dakurah2025topologically}.

\subsection{Difference in the efficiency of information flow between networks}
The organization of the shortest paths in networks have a direct influence on the flow of information around the network. Networks with shortest paths that exhibits stronger edge weights, involving fewer links with lower node degree will have a more efficient transfer of information along its communication paths. We can quantify how two networks differ in the transfer of information using the information measure of the shortest paths introduced in previous sections. Let $\mathbb{K}_1, \mathbb{K}_2$ be a collection of networks in  two groups. We assume all the networks in the two groups have the same set of nodes. The networks in the two groups will have the same node set $\mathcal{V}$ but with different information defined on the edges.  Let $\bar{I}_1(s, t)$ and $\bar{I}_2(s, t)$ be the average information of the shortest path linking $s$ and $t$ for the the networks in $\mathbb{K}_1$ and $\mathbb{K}_2$ respectively. Since we're interested in detecting which paths the two networks differ the most interms of the information measure, we consider the inverse of the information average measures $\tau_1(s, t) = \bar{I}_1^{-1}(s, t)$, $\tau_2(s, t) = \bar{I}_2^{-1}(s, t)$. Then the proposed test statistic for testing the difference in the efficiency of information flow  between the two groups  is given by:

\begin{equation}
    \mathcal{I}(\mathbb{K}_1, \mathbb{K}_2) = \max\limits_{\substack{{s, t \in \mathcal{V}} \\ s < t}} |\tau_1(s, t) - \tau_2(s, t)|.
    \label{eqn:test-statistic-info}
\end{equation}
If the shortest path connecting the source node $s$ and target $t$ differs between the two groups of networks, then $\mathcal{I}(\mathbb{K}_1, \mathbb{K}_2)$ should be large. It will be interesting to observe whether the paths contributing to this difference belongs to the spanning trees. The statistical significance is determined using the permutation test.

\section{Validation}
We perform a series of validation studies that demonstrate the superior performance of our method to existing methods in the literature. 

\subsection{Study 1: Robustness of the spectral averaging}
Constructing the group level tree from the average network introduces bias from individual networks with outlying connectivities \cite{ciftci2011minimum,van2018minimum,dakurah2024subsequence}. That's, the average tree in this case will largely reflect the topology of the networks with the outlying connectivities, wherein a significant proportion of the edges in the average MST will be associated with the MST of the network with outlying connections. Here, an outlying connection in a network is quantified as follows. Given a collection of edge weights of a network, we term an edge weight to be outlying (or an outlying connection) if it exceeds in absolute terms, twice the difference between the first quartile and the third quartile of the collection.  We show that the spectral averaging minimizes this outlying effect. 

{\em Networks construction}
We generate two group of networks $\mathbb{K}_1$ and $\mathbb{K}_2$, with each group having exactly 10 networks and each network has $m$ nodes. The edge weights of the networks in $\mathbb{K}_1$ were drawn randomly from a uniform distribution on the interval $(0, 1)$.  To induce outlying connections in the networks in $\mathbb{K}_2$, we initially draw the edge weights of all the networks in $\mathbb{K}_2$ randomly from a uniform distribution on the interval $(0, 0.3)$. Then for each network, we randomly select $m-1$ edges and set their edge weights to a random draw from a uniform distribution on the interval $(0.7, 0.9)$. This construction induces not less than $m-1$ outlying connections in the network. Then $\mathbb{K}_1$ is the group in which the networks do not have outlying connections whiles $\mathbb{K}_2$ is the group in which the networks have outlying connections.
Our interest is in computing the average tree of the combination of the networks in $\mathbb{K}_1$ and $\mathbb{K}_2$. There are four different combinations: $\{ (\mathbb{K}_i, \mathbb{K}_j) : 1 \le i, j \le 2 \}$. We also denote the corresponding collection of the MSTs as $\{(\mathbb{T}_i, \mathbb{T}_j) : 1 \le i, j \le 2 \}$. We can compute the average MST in two formats. One based on computing the average of all the network connectivity matrices in the combined groups. Subsequently, the average tree is computed based on this average connectivity matrix. Henceforth this procedure will be referred to as the ordinary averaging method. The second option is based on our  proposed spectral averaging method.
\begin{figure}[ht]
	\centering
	\includegraphics[width =\textwidth]{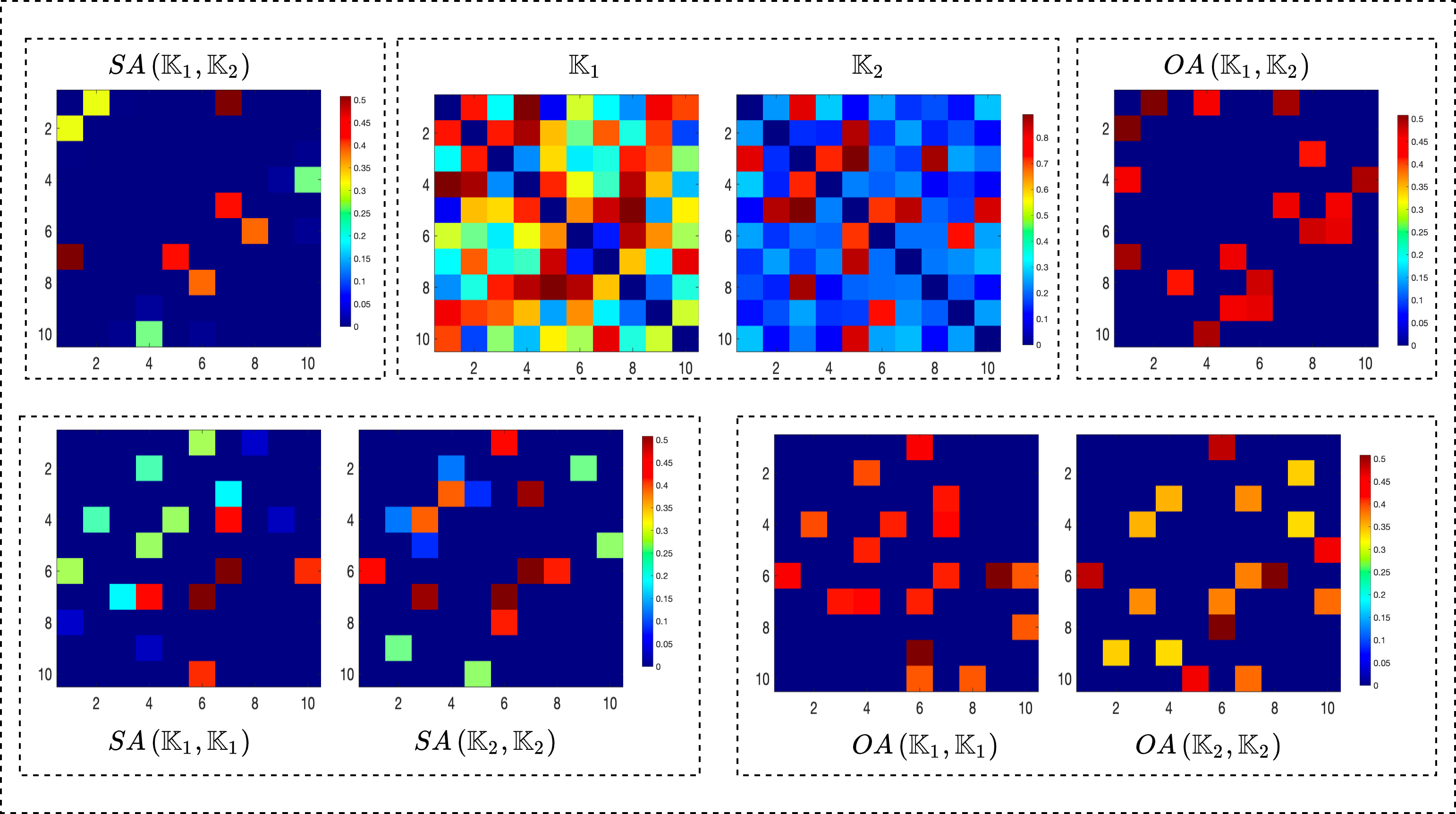}
	\caption{Illustration of the average tree computation for a combination of the two group of networks $\mathbb{K}_1$ and $\mathbb{K}_2$, where each group has 10 networks with 10 nodes each. $SA(.,.)$ denotes the spectral average tree network, $OA(.,.)$ denote the ordinary average tree network.}
	\label{fig:outliers-networks}
\end{figure}
Figure \ref{fig:outliers-networks} illustrates an example of the average tree computation for a combination of the two group of networks $\mathbb{K}_1$ and $\mathbb{K}_2$ where each group has 10 networks and each network is fully connected with 10 nodes. $\mathbb{K}_2$ contain networks with outlying connections.

{\em Comparison procedure} We hypothesize that, the spectral averaging dampens the effect of outlying connections in individual networks. To develop a numerical procedure for quantifying this, we proceed as follows. Let $\bar{T}_O$ and $\bar{T}_S$ be the average tree based on the ordinary average and the spectral averaging method respectively. First for the ordinary averaging, let $\rho(T_i, \bar{T}_O)$ be the proportion of edges in the tree $T_i$ that are also in the average tree $\bar{T}_O$, we then have the following statistic
\begin{equation}
\Gamma(\mathbb{T}_i, \mathbb{T}_j) = \sum_{T_i \in \mathbb{T}_i, T_j \in \mathbb{T}_j} |\rho(T_i, \bar{T}_O) - \rho(T_j, \bar{T}_O)|.
\label{eqn:test-stat-outliers}
\end{equation}
Similarly, for the case of spectral averaging, we replace $\bar{T}_O$ with $\bar{T}_S$. Under the null hypothesis assumption that there are no outlying connections in the networks of the two groups been combined, $\rho(T_i, \bar{T}_O) \approx \rho(T_j, \bar{T}_O)$, hence $\Gamma(\mathbb{T}_i, \mathbb{T}_j)$ should be close to $0$. The deviation from $0$ indicates the effect of outlying connections. The statistical significance will be determined using the permutation test.

For each combination of groups $\mathbb{K}_1$ and $\mathbb{K}_2$, the permutation test with $100000$ permutations were run and the process repeated 10 times. The average p-value and it's standard deviation are reported in Table \ref{tab:SimulationStudy1-outliers}. We varied the node size to see the influence of the network size on the two methods. When comparing networks of the same group (first four rows), both the spectral and ordinary averaging method correctly indicates there's no outlying effect in the computation of the average MST (high p-values).
\begin{table}[ht!]
\caption{The performance results summarized as average p-values for comparing the two groups of networks. Larger values are preferred regardless of the comparison group. For comparing the same group of networks (first four rows), no outlier effect should be detected regardless of the method. When comparing groups where one has networks with outlying connections (last two rows), a method that reduces the influence of outliers will conclude no outlier effect (spectral average), otherwise it will indicate the influence of outliers (ordinary average, grayed row).}
\label{tab:SimulationStudy1-outliers}
\setlength{\tabcolsep}{4.5pt}
\renewcommand{\arraystretch}{1.5}
\centering
\begin{tabular}{cc|ccccc}
    Group & Method & 10 nodes & 25 nodes & 50 nodes & 100 nodes \\ \hline
  \multirow{2}{*}{$\mathbb{K}_1$ vs. $\mathbb{K}_1$} & Ordinary Average & $0.52 \pm 0.07$ & $0.54 \pm 0.06$ & $0.50 \pm 0.01$ & $0.52 \pm 0.02$\\ 
   & Spectral Average & $0.55 \pm 0.04$ & $0.56 \pm 0.02$ & $0.52 \pm 0.02$ & $0.64 \pm 0.06$ \\ 
  
   \multirow{2}{*}{$\mathbb{K}_2$ vs. $\mathbb{K}_2$} & Ordinary Average & $0.53 \pm 0.04$ & $0.54 \pm 0.02$ & $0.52 \pm 0.02$ & $0.51 \pm 0.01$ \\
   & Spectral Average & $0.54 \pm 0.06$ & $0.53 \pm 0.02$ & $0.51 \pm 0.03$ & $0.51 \pm 0.01 $\\ \hline
  
  \rowcolor{Gray}
  \cellcolor{white} \multirow{2}{*}{$\mathbb{K}_1$ vs. $\mathbb{K}_2$} & \cellcolor{white}  Ordinary Average & $0.07 \pm 0.11$ & $0.04 \pm 0.05$ & $0.01 \pm 0.11$ & $0.01 \pm 0.05$ \\
   & Spectral Average & $0.57 \pm 0.29$ & $0.36 \pm 0.24$ & $0.42 \pm 0.25$ & $0.41 \pm 0.22$\\ 
 \hline \hline
\end{tabular}
\end{table}
 When comparing $\mathbb{K}_1$ and $\mathbb{K}_2$ however, we notice since $\mathbb{K}_2$ contains networks with outlying connections, a good method will reduce the effect of these outlying connections and hence the test will conclude that there's no outlier influence in computing the average MST. The ordinary average method led to the conclusion that there's an influence of outliers in computing the average MST (grayed row) whiles the spectral average method indicates there is no influence of outliers in the average MST. This study confirms our assertion that the spectral averaging method is robust to outliers in the individual network connectivities.

\section{Discussion and Conclusion}

This work addresses a fundamental challenge in network neuroscience: how to compute unbiased, computationally efficient representations of group-level brain network topologies. We introduced a novel spectral representation of maximum spanning trees using the spanning tree basis, enabling the first principled approach to averaging collections of trees while preserving their essential topological properties. By leveraging the Hodge Laplacian and its eigendecomposition, we demonstrated that trees, which lack natural vector space representations, can be expressed as linear combinations of fundamental cycle bases, enabling mathematical operations on trees that were previously intractable. The spectral averaging method demonstrates clear advantages over existing approaches. Traditional methods that construct group-level trees from averaged connectivity matrices are inherently biased toward subjects with outlying connection strengths, potentially distorting the representation of typical network organization. Our validation studies confirmed that spectral averaging substantially reduces this bias, maintaining robustness even when individual networks contain significant outlying connections. The consistently high p-values $0.36$ to $0.57$ when comparing groups with and without outliers, compared to the ordinary averaging method's low p-values $0.01$ to$0.07$, demonstrate the superior robustness of our approach. Bootstrap-based methods, while theoretically sound, suffer from computational complexity that grows prohibitively with network size, whereas our spectral approach provides a deterministic, closed-form solution that scales efficiently.

The introduction of path information as a measure of communication efficiency represents a novel contribution to network comparison methodology. By quantifying the accessibility of shortest paths based on edge weights and node degrees, this measure captures aspects of network organization that complement traditional graph-theoretic metrics. The path information measure is particularly well-suited to functional brain networks, where edge weights represent correlation strengths. Networks with lower path information exhibit more concentrated information flow along major communication highways, potentially indicating more efficient or specialized functional organization. The maximum test statistic approach allows for localization of differences in information flow efficiency between network groups, identifying specific source-target pairs where communication pathways differ most substantially. The scale-invariant properties of MSTs make them particularly valuable for comparing brain networks across subjects with varying overall connectivity strengths, a persistent challenge in functional connectivity analysis. Our spectral framework maintains this scale invariance while enabling group-level inference, providing a principled foundation for studying sex differences, developmental changes, or pathological alterations in brain network organization. The demonstration that average MSTs overlap substantially with the union of shortest paths in functional networks supports their validity as backbone representations, suggesting that the edges retained in average MSTs correspond to the most critical pathways for information integration across networks.

This work presents a unified framework for computing unbiased averages of network topologies using the spanning tree basis representation. This spectral approach addresses fundamental limitations of existing methods by providing computational efficiency, robustness to outliers, and mathematical rigor. The spanning tree basis transforms the challenging problem of averaging trees into a tractable linear algebra problem, enabling standard statistical inference procedures on tree topologies. Combined with the path information measure for quantifying communication efficiency, this framework provides powerful new tools for comparing functional brain network organization across groups. By enabling group-level tree analysis with minimal bias and computational overhead, this work will facilitate large-scale studies of brain network organization and its alterations across populations, developmental stages, and disease states, advancing our ability to understand the organizational principles of complex brain networks and their relevance to human cognition and behavior, and will be explored in future work.

\bibliographystyle{ieeetr}
\bibliography{references}

@article{topaz2015topological,
  title={Topological data analysis of biological aggregation models},
  author={Topaz, Chad M and Ziegelmeier, Lori and Halverson, Tom},
  journal={PloS one},
  volume={10},
  number={5},
  pages={e0126383},
  year={2015},
  publisher={Public Library of Science San Francisco, CA USA}
}

@article{bullmore2009complex,
  title={Complex brain networks: graph theoretical analysis of structural and functional systems},
  author={Bullmore, Ed and Sporns, Olaf},
  journal={Nature reviews neuroscience},
  volume={10},
  number={3},
  pages={186--198},
  year={2009},
  publisher={Nature Publishing Group}
}

@article{torres2021and,
  title={The why, how, and when of representations for complex systems},
  author={Torres, Leo and Blevins, Ann S and Bassett, Danielle and Eliassi-Rad, Tina},
  journal={SIAM Review},
  volume={63},
  number={3},
  pages={435--485},
  year={2021},
  publisher={SIAM}
}

@article{park2013structural,
  title={Structural and functional brain networks: from connections to cognition},
  author={Park, Hae-Jeong and Friston, Karl},
  journal={Science},
  volume={342},
  number={6158},
  year={2013},
  publisher={American Association for the Advancement of Science}
}

@inproceedings{lee2014hole,
  title={Hole detection in metabolic connectivity of Alzheimer’s disease using k- Laplacian},
  author={Lee, Hyekyoung and Chung, Moo K and Kang, Hyejin and Lee, Dong Soo},
  booktitle={International Conference on Medical Image Computing and Computer-Assisted Intervention},
  pages={297--304},
  year={2014},
  organization={Springer}
}

@article{tzourio2002automated,
  title={Automated anatomical labeling of activations in SPM using a macroscopic anatomical parcellation of the MNI MRI single-subject brain},
  author={Tzourio-Mazoyer, Nathalie and Landeau, Brigitte and Papathanassiou, Dimitri and Crivello, Fabrice and Etard, Olivier and Delcroix, Nicolas and Mazoyer, Bernard and Joliot, Marc},
  journal={Neuroimage},
  volume={15},
  number={1},
  pages={273--289},
  year={2002},
  publisher={Elsevier}
}

@book{bondy1976graph,
  title={Graph theory with applications},
  author={Bondy, John Adrian and Murty, Uppaluri Siva Ramachandra and others},
  volume={290},
  year={1976},
  publisher={Macmillan London}
}

@inproceedings{smith2015comparison,
  title={Comparison of network analysis approaches on EEG connectivity in beta during Visual Short-term Memory binding tasks},
  author={Smith, Keith and Azami, Hamed and Escudero, Javier and Parra, Mario A and Starr, John M},
  booktitle={2015 37th Annual International Conference of the IEEE Engineering in Medicine and Biology Society (EMBC)},
  pages={2207--2210},
  year={2015},
  organization={IEEE}
}

@inproceedings{benabdallah2020analysis,
  title={Analysis of the over-connectivity in autistic brains using the maximum spanning tree: application on the multi-site and heterogeneous ABIDE dataset},
  author={Benabdallah, Fatima Zahra and El Maliani, Ahmed Drissi and Lotfi, Dounia and El Hassouni, Mohammed},
  booktitle={2020 8th International Conference on Wireless Networks and Mobile Communications (WINCOM)},
  pages={1--7},
  year={2020},
  organization={IEEE}
}

@inproceedings{wang2021multilayer,
  title={A Multilayer Maximum Spanning Tree Kernel For Brain Networks},
  author={Wang, Xiaoxin and Wen, Xuyun and Ma, Kai and Zhang, Daoqiang},
  booktitle={2021 IEEE 18th International Symposium on Biomedical Imaging (ISBI)},
  pages={1582--1585},
  year={2021},
  organization={IEEE}
}

@article{aigner1999proofs,
  title={Proofs from the Book},
  author={Aigner, Martin and Ziegler, G{\"u}nter M},
  journal={Berlin. Germany},
  year={1999},
  publisher={Springer}
}

@article{kruskal1956shortest,
  title={On the shortest spanning subtree of a graph and the traveling salesman problem},
  author={Kruskal, Joseph B},
  journal={Proceedings of the American Mathematical society},
  volume={7},
  number={1},
  pages={48--50},
  year={1956},
  publisher={JSTOR}
}

@article{goldenberg2015use,
  title={The use of functional and effective connectivity techniques to understand the developing brain},
  author={Goldenberg, Diane and Galv{\'a}n, Adriana},
  journal={Developmental cognitive neuroscience},
  volume={12},
  pages={155--164},
  year={2015},
  publisher={Elsevier}
}

@inproceedings{huang2019dynamic,
  title={Dynamic functional connectivity using heat kernel},
  author={Huang, Shih-Gu and Chung, Moo K and Carroll, Ian C and Goldsmith, H Hill},
  booktitle={2019 IEEE Data Science Workshop (DSW)},
  pages={222--226},
  year={2019},
  organization={IEEE}
}

@inproceedings{songdechakraiwut2020dynamic,
  title={Dynamic topological data analysis for functional brain signals},
  author={Songdechakraiwut, Tananun and Chung, Moo K},
  booktitle={2020 IEEE 17th International Symposium on Biomedical Imaging Workshops (ISBI Workshops)},
  pages={1--4},
  year={2020},
  organization={IEEE}
}

@article{honey2007network,
  title={Network structure of cerebral cortex shapes functional connectivity on multiple time scales},
  author={Honey, Christopher J and K{\"o}tter, Rolf and Breakspear, Michael and Sporns, Olaf},
  journal={Proceedings of the National Academy of Sciences},
  volume={104},
  number={24},
  pages={10240--10245},
  year={2007},
  publisher={National Acad Sciences}
}

@article{dimitriadis2017topological,
  title={Topological filtering of dynamic functional brain networks unfolds informative chronnectomics: a novel data-driven thresholding scheme based on orthogonal minimal spanning trees (OMSTs)},
  author={Dimitriadis, Stavros I and Salis, Christos and Tarnanas, Ioannis and Linden, David E},
  journal={Frontiers in neuroinformatics},
  volume={11},
  pages={28},
  year={2017},
  publisher={Frontiers}
}

@article{farahani2019application,
  title={Application of graph theory for identifying connectivity patterns in human brain networks: a systematic review},
  author={Farahani, Farzad V and Karwowski, Waldemar and Lighthall, Nichole R},
  journal={frontiers in Neuroscience},
  volume={13},
  pages={585},
  year={2019},
  publisher={Frontiers}
}

@article{wang2010graph,
  title={Graph-based network analysis of resting-state functional MRI},
  author={Wang, Jinhui and Zuo, Xinian and He, Yong},
  journal={Frontiers in systems neuroscience},
  volume={4},
  pages={16},
  year={2010},
  publisher={Frontiers}
}

@inproceedings{lee2012weighted,
  title={Weighted functional brain network modeling via network filtration},
  author={Lee, Hyekyoung and Kang, Hyejin and Chung, Moo K and Kim, Bung-Nyun and Lee, Dong Soo},
  booktitle={NIPS Workshop on Algebraic Topology and Machine Learning},
  volume={3},
  year={2012},
  organization={Citeseer}
}

@article{van2018minimum,
  title={Minimum spanning tree analysis of the human connectome},
  author={van Dellen, Edwin and Sommer, Iris E and Bohlken, Marc M and Tewarie, Prejaas and Draaisma, Laurijn and Zalesky, Andrew and Di Biase, Maria and Brown, Jesse A and Douw, Linda and Otte, Willem M and others},
  journal={Human brain mapping},
  volume={39},
  number={6},
  pages={2455--2471},
  year={2018},
  publisher={Wiley Online Library}
}

@article{tewarie2015minimum,
  title={The minimum spanning tree: an unbiased method for brain network analysis},
  author={Tewarie, Prejaas and van Dellen, Edwin and Hillebrand, Arjan and Stam, Cornelis J},
  journal={Neuroimage},
  volume={104},
  pages={177--188},
  year={2015},
  publisher={Elsevier}
}

@book{kaczynski2004computational,
  title={Computational homology},
  author={Kaczynski, Tomasz and Mischaikow, Konstantin Michael and Mrozek, Marian},
  volume={3},
  year={2004},
  publisher={Springer}
}

@article{bach2013learning,
  title={Learning with submodular functions: A convex optimization perspective},
  author={Bach, Francis and others},
  journal={Foundations and Trends{\textregistered} in Machine Learning},
  volume={6},
  number={2-3},
  pages={145--373},
  year={2013},
  publisher={Now Publishers, Inc.}
}

@article{jost2021discrete,
  title={Discrete-to-Continuous Extensions: Lov$\backslash$'asz extension, optimizations and eigenvalue problems},
  author={Jost, J{\"u}rgen and Zhang, Dong},
  journal={arXiv preprint arXiv:2106.03189},
  year={2021}
}

@article{barbarossa2020topological,
  title={Topological signal processing over simplicial complexes},
  author={Barbarossa, Sergio and Sardellitti, Stefania},
  journal={IEEE Transactions on Signal Processing},
  volume={68},
  pages={2992--3007},
  year={2020},
  publisher={IEEE}
}

@article{von2007tutorial,
  title={A tutorial on spectral clustering},
  author={Von Luxburg, Ulrike},
  journal={Statistics and computing},
  volume={17},
  number={4},
  pages={395--416},
  year={2007},
  publisher={Springer}
}

@article{ortega2018graph,
  title={Graph signal processing: Overview, challenges, and applications},
  author={Ortega, Antonio and Frossard, Pascal and Kova{\v{c}}evi{\'c}, Jelena and Moura, Jos{\'e} MF and Vandergheynst, Pierre},
  journal={Proceedings of the IEEE},
  volume={106},
  number={5},
  pages={808--828},
  year={2018},
  publisher={IEEE}
}

@article{dancso2016construction,
  title={A construction of the graphic matroid from the lattice of integer flows},
  author={Dancso, Zsuzsanna and Garoufalidis, Stavros},
  journal={arXiv preprint arXiv:1611.06282},
  year={2016}
}

@inproceedings{zach2010disambiguating,
  title={Disambiguating visual relations using loop constraints},
  author={Zach, Christopher and Klopschitz, Manfred and Pollefeys, Marc},
  booktitle={2010 IEEE Computer Society Conference on Computer Vision and Pattern Recognition},
  pages={1426--1433},
  year={2010},
  organization={IEEE}
}

@article{barr1979enhancements,
  title={Enhancements of spanning tree labelling procedures for network optimization},
  author={Barr, Richard and Glover, Fred and Klingman, Darwin},
  journal={INFOR: Information Systems and Operational Research},
  volume={17},
  number={1},
  pages={16--34},
  year={1979},
  publisher={Taylor \& Francis}
}

@article{umeyama1988eigendecomposition,
  title={An eigendecomposition approach to weighted graph matching problems},
  author={Umeyama, Shinji},
  journal={IEEE transactions on pattern analysis and machine intelligence},
  volume={10},
  number={5},
  pages={695--703},
  year={1988},
  publisher={IEEE}
}

@article{demetci2020gromov,
  title={Gromov-Wasserstein optimal transport to align single-cell multi-omics data},
  author={Demetci, Pinar and Santorella, Rebecca and Sandstede, Bj{\"o}rn and Noble, William Stafford and Singh, Ritambhara},
  journal={BioRxiv},
  year={2020},
  publisher={Cold Spring Harbor Laboratory}
}

@article{maretic2020wasserstein,
  title={Wasserstein-based graph alignment},
  author={Maretic, Hermina Petric and Gheche, Mireille El and Minder, Matthias and Chierchia, Giovanni and Frossard, Pascal},
  journal={arXiv preprint arXiv:2003.06048},
  year={2020}
}

@inproceedings{white2006mixing,
  title={Mixing spectral representations of graphs},
  author={White, David and Wilson, Richard C},
  booktitle={18th International Conference on Pattern Recognition (ICPR'06)},
  volume={4},
  pages={140--144},
  year={2006},
  organization={IEEE}
}

@article{stam2014trees,
  title={The trees and the forest: characterization of complex brain networks with minimum spanning trees},
  author={Stam, CJ and Tewarie, P and Van Dellen, E and Van Straaten, ECW and Hillebrand, A and Van Mieghem, P},
  journal={International Journal of Psychophysiology},
  volume={92},
  number={3},
  pages={129--138},
  year={2014},
  publisher={Elsevier}
}

@article{ciftci2011minimum,
  title={Minimum spanning tree reflects the alterations of the default mode network during Alzheimer’s disease},
  author={Ciftci, KORAY},
  journal={Annals of biomedical engineering},
  volume={39},
  number={5},
  pages={1493--1504},
  year={2011},
  publisher={Springer}
}

@article{engels2015declining,
  title={Declining functional connectivity and changing hub locations in Alzheimer’s disease: an EEG study},
  author={Engels, Marjolein and Stam, Cornelis J and van der Flier, Wiesje M and Scheltens, Philip and de Waal, Hanneke and van Straaten, Elisabeth CW},
  journal={BMC neurology},
  volume={15},
  number={1},
  pages={1--8},
  year={2015},
  publisher={Springer}
}

@article{otte2015aging,
  title={Aging alterations in whole-brain networks during adulthood mapped with the minimum spanning tree indices: the interplay of density, connectivity cost and life-time trajectory},
  author={Otte, Willem M and Van Diessen, Eric and Paul, Subhadip and Ramaswamy, Rajiv and Rallabandi, VP Subramanyam and Stam, Cornelis J and Roy, Prasun K},
  journal={Neuroimage},
  volume={109},
  pages={171--189},
  year={2015},
  publisher={Elsevier}
}

@article{lee2006classification,
  title={Classification of epilepsy types through global network analysis of scalp electroencephalograms},
  author={Lee, UnCheol and Kim, Seunghwan and Jung, Ki-Young},
  journal={Physical Review E},
  volume={73},
  number={4},
  pages={041920},
  year={2006},
  publisher={APS}
}

@article{de2013estimating,
  title={Estimating false positives and negatives in brain networks},
  author={de Reus, Marcel A and van den Heuvel, Martijn P},
  journal={Neuroimage},
  volume={70},
  pages={402--409},
  year={2013},
  publisher={Elsevier}
}

@book{mieghem2006data,
  title={Data communications networking},
  author={Mieghem, Piet van},
  year={2006},
  publisher={Purdue University Press}
}

@book{van2010graph,
  title={Graph spectra for complex networks},
  author={Van Mieghem, Piet},
  year={2010},
  publisher={Cambridge University Press}
}

@article{van2008observable,
  title={The observable part of a network},
  author={Van Mieghem, Piet and Wang, Huijuan},
  journal={IEEE/ACM Transactions on Networking},
  volume={17},
  number={1},
  pages={93--105},
  year={2008},
  publisher={IEEE}
}

@article{sneppen2005hide,
  title={Hide-and-seek on complex networks},
  author={Sneppen, Kim and Trusina, Ala and Rosvall, Martin},
  journal={EPL (Europhysics Letters)},
  volume={69},
  number={5},
  pages={853},
  year={2005},
  publisher={IOP Publishing}
}

@article{trusina2005communication,
  title={Communication boundaries in networks},
  author={Trusina, Ala and Rosvall, Martin and Sneppen, Kim},
  journal={Physical review letters},
  volume={94},
  number={23},
  pages={238701},
  year={2005},
  publisher={APS}
}

@article{latora2001efficient,
  title={Efficient behavior of small-world networks},
  author={Latora, Vito and Marchiori, Massimo},
  journal={Physical review letters},
  volume={87},
  number={19},
  pages={198701},
  year={2001},
  publisher={APS}
}

@article{meier2015union,
  title={The union of shortest path trees of functional brain networks},
  author={Meier, Jil and Tewarie, Prejaas and Van Mieghem, Piet},
  journal={Brain Connectivity},
  volume={5},
  number={9},
  pages={575--581},
  year={2015},
  publisher={Mary Ann Liebert, Inc. 140 Huguenot Street, 3rd Floor New Rochelle, NY 10801 USA}
}

@article{dakurah2025discrete,
  title={Discrete Heat Kernels on Simplicial Complexes and Its Application to Functional Brain Networks},
  author={Dakurah, Sixtus},
  journal={arXiv preprint arXiv:2509.16908},
  year={2025}
}

@article{dakurahregistration,
  title={Registration and Joint Identification of Cycles in Brain Networks},
  author={Dakurah, Sixtus and Chung, Moo K}
}

@inproceedings{dakurah2022modelling,
  title={Modelling cycles in brain networks with the Hodge Laplacian},
  author={Dakurah, Sixtus and Anand, D Vijay and Chen, Zijian and Chung, Moo K},
  booktitle={International Conference on Medical Image Computing and Computer-Assisted Intervention},
  pages={326--335},
  year={2022},
  organization={Springer}
}

@article{dakurah2025maxtda,
  title={MaxTDA: Robust Statistical Inference for Maximal Persistence in Topological Data Analysis},
  author={Dakurah, Sixtus and Cisewski-Kehe, Jessi},
  journal={arXiv preprint arXiv:2504.03897},
  year={2025}
}

@book{dakurah2025robust,
  title={Robust Statistical Methods for Topological Data Analysis of Time Series},
  author={Dakurah, Sixtus},
  year={2025},
  publisher={The University of Wisconsin-Madison}
}

@article{dakurah2025topologically,
  title={Topologically Invariant Permutation Test},
  author={Dakurah, Sixtus},
  journal={arXiv preprint arXiv:2511.06153},
  year={2025}
}

@article{dakurah2024subsequence,
  title={A subsequence approach to topological data analysis for irregularly-spaced time series},
  author={Dakurah, Sixtus and Cisewski-Kehe, Jessi},
  journal={arXiv preprint arXiv:2410.13723},
  year={2024}
}

@article{dakurah2025brain,
  title={Brain Networks Flow-Topology via Variance Minimization in the Wasserstein Space},
  author={Dakurah, Sixtus},
  journal={arXiv preprint arXiv:2511.12990},
  year={2025}
}
\end{document}